\documentclass[aps,prl,superscriptaddress,floatfix,twocolumn,showpacs]{revtex4}

\pdfoutput=1

\usepackage{amsmath,dcolumn,bm,amsthm,amsfonts,amssymb,tabularx,subfigure}
\usepackage[colorlinks=true,urlcolor=blue,citecolor=blue,linkcolor=blue]{hyperref}
\usepackage{mathrsfs}
\usepackage{times,bbold,dsfont}
\usepackage{graphicx,graphics,color}
\usepackage{mathtools}

\newcommand{\beq}{\begin{equation}}
\newcommand{\eeq}{\end{equation}}
\newcommand{\bea}{\begin{eqnarray}}
\newcommand{\eea}{\end{eqnarray}}
\newcommand{\bal}{\begin{align}}
\newcommand{\eal}{\end{align}}
\newcommand{\rmI}{\textrm{I}}
\newcommand{\rmII}{\textrm{II}}
\newcommand{\tpsi}{\tilde{\psi}}
\newcommand{\Pv}{{\sf P_v}}
\newcommand{\Ls}{{\sf L_s}}
\newcommand{\mz}{\mathbb{Z}}
\newcommand{\imi}{i}

\begin{document}

\title{Flux-Stabilized Majorana Zero Modes in Coupled One-Dimensional Fermi Wires}

\author{Chun~Chen}
\email[Corresponding author. \\]{cchen@physics.umn.edu}
\affiliation{School of Physics and Astronomy, University of Minnesota, Minneapolis, Minnesota 55455, USA}

\author{Wei~Yan}
\affiliation{Department of Physics and State Key Laboratory of Surface Physics, Fudan University, Shanghai 200433, China}
\affiliation{Collaborative Innovation Center of Advanced Microstructures, Nanjing 210093, China}

\author{C.~S.~Ting}
\affiliation{Texas Center for Superconductivity and Department of Physics, University of Houston, Houston, Texas 77204, USA}

\author{Yan~Chen}
\affiliation{Department of Physics and State Key Laboratory of Surface Physics, Fudan University, Shanghai 200433, China}
\affiliation{Collaborative Innovation Center of Advanced Microstructures, Nanjing 210093, China}

\author{F.~J.~Burnell}
\affiliation{School of Physics and Astronomy, University of Minnesota, Minneapolis, Minnesota 55455, USA}

\date{\today}

\begin{abstract}

One promising avenue to study one-dimensional ($1$D) topological phases is to realize them in synthetic materials such as cold atomic gases. Intriguingly, it is possible to realize Majorana boundary modes in a $1$D number-conserving system consisting of two fermionic chains coupled only by pair-hopping processes \cite{Kraus}. It is commonly believed that significant interchain single-particle tunneling necessarily destroys these Majorana modes, as it spoils the $\mathbb{Z}_2$ fermion parity symmetry that protects them. In this Letter, we present a new mechanism to overcome this obstacle, by piercing a (synthetic) magnetic $\pi$-flux through each plaquette of the Fermi ladder. Using bosonization, we show that in this case there exists an exact leg-interchange symmetry that is robust to interchain hopping, and acts as fermion parity at long wavelengths. We utilize density matrix renormalization group and exact diagonalization to verify that the resulting model exhibits Majorana boundary modes up to large single-particle tunnelings, comparable to the intrachain hopping strength. Our work highlights the unusual impacts of different topologically trivial band structures on these interaction-driven topological phases, and identifies a distinct route to stabilizing Majorana boundary modes in $1$D fermionic ladders.

\end{abstract}

\pacs{67.85.$-$d, 71.10.Pm, 03.67.Lx, 74.90.$+$n}

\maketitle

The classification of topological phases in one dimension \cite{Pollmann,FidkowskiKitaevClass,ChenGuWen} revealed an intriguing array of possible new states of matter, with a variety of types of protected gapless boundary modes. Of these, one class that has generated considerable excitement recently is the one-dimensional ($1$D) topological superconductors first described by Ref.~\cite{Kitaev}. These have protected boundary Majorana zero modes, which harbor non-Abelian statistics \cite{AliceaBraiding,SauPRB,Hassler} and hence are promising candidates for topological quantum computing \cite{NayakReview,SarmaReview}.

In practise, to obtain long-range superconducting order in $1$D systems requires inducing superconductivity via coupling to a $3$D bulk superconductor \cite{LFu,Fu,AliceaPRB,JDSau,Lutchyn,Oreg,Vazifeh,Klinovaja,Braunecker,Alicea,Clarke}; experimental progress in this direction has been made in several distinct solid-state systems \cite{Mourik,Das,Deng,NPerge,Albrecht}. Interestingly, however, it is also possible to host topological boundary modes in truly $1$D platforms, in spite of the fact that these systems do not support long-range superconducting order, and are in fact gapless \cite{Fidkowski,Sau,Kraus,Cheng,Lang,Iemini}. This opens up the possibility of studying $1$D fermionic topological phases in synthetic materials such as cold atomic gases \cite{Zhang,Sato,Chen,Qu,WZhang,Liu,BlochRMP,Goldman}, offering an attractive architecture with increased tunability.

One concrete model in this category was proposed by Ref.~\cite{Kraus}, who showed numerically that even starting from a lattice Hamiltonian with a topologically trivial band structure, a regime bearing the hallmarks of Majorana boundary modes can be accessed in an atomic two-leg ladder by introducing an interleg pair-hopping interaction. These boundary modes are protected by the conserved fermion parity of \emph{one} of the wires, and are therefore robust provided that the single-particle interleg tunneling ($t_{\perp}$), which breaks this symmetry explicitly, is sufficiently small. A number of proposals \cite{Kraus,Cheng,Lang,Iemini} for suppressing $t_{\perp}$ such that this regime may be experimentally realized ensued.

In this paper we propose a distinct route to overcome this obstacle: We begin with a different band structure, in which each plaquette of the ladder has a flux of $\pi$. We show that with pair hopping this model also has an interacting topological regime hosting Majorana boundary modes---which in this case are protected by a symmetry preserved even in the presence of finite $t_{\perp}$. We bolster our theory with numerical evidence of Majorana boundary modes over a wide range of $t_{\perp}$. Our findings thus furnish an appealing mechanism to engineer topological boundary modes in a particle-conserving and strongly interacting system without the need to fine-tune single-particle tunneling.

{\em Fermionic flux ladder model.}---Motivated by these considerations, we study an interacting two-leg ladder model of spinless fermions in a perpendicular magnetic field described by the following number-conserving Hamiltonian,
\begin{align}
H=&\ H_K+H_W, \label{modelhamiltonian} \\
H_K=&-\!\sum^{L-2}_{n=0}[(t_{\parallel} e^{i\frac{\phi}{2}} c^{\dagger}_{n,0}c_{n+1,0}\!+\!t_{\parallel} e^{-i\frac{\phi}{2}} c^{\dagger}_{n,1}c_{n+1,1})\!+\!\textrm{H.c.}] \nonumber \\
&-\!\sum^{L-1}_{n=0}(t_{\perp}c^{\dagger}_{n,0}c_{n,1}\!+\!\textrm{H.c.}), \label{modelhamiltonianH_K} \\
H_W=&+\!\sum^{L-2}_{n=0}(Wc^{\dagger}_{n,0}c^{\dagger}_{n+1,0}c_{n,1}c_{n+1,1}\!+\!\textrm{H.c.}),
\end{align}
where $c^{(\dagger)}_{n,\ell}$ is the fermionic annihilation $($creation$)$ operator at rung $n$ on the leg $\ell\!=\!0,1$. The intraleg and interleg single-particle tunneling strengths are $t_{\parallel}$ and $t_{\perp}$, respectively, and $W$ (which we take to be negative throughout this work) is the pair-hopping strength. Two essential ingredients of the above model are the synthetic Peierls phase $\phi\!\in\![0,\pi]$ per plaquette, and the interchain pair-hopping interaction $H_W$. Previous works have demonstrated the existence of Majorana boundary modes in this Hamiltonian at $\phi\!=\!0$ and $t_{\perp}\!=\!0$ based on a preserved fermion number parity ${\sf P}_\ell\!\coloneqq\!(-1)^{N_{\ell}}$, where $N_{\ell}$ is the particle-number operator of a single leg $\ell$ \cite{Cheng,Kraus,Lang,Iemini}. Here we will show that when $\phi\!=\!\pi$, these boundary modes persist up to $t_{\perp}$ of order $t_{\parallel}$. Note that without pair hopping, the bare band $H_K$ in (\ref{modelhamiltonianH_K}) is topologically trivial, so that the model (\ref{modelhamiltonian}) requires interactions to realize the topological regime. This is necessarily the case for isolated $1$D systems, where the total fermion number is conserved---in contrast to models based on e.g., Kitaev or spin-orbit-coupled wires \cite{Stoudenmire,Chan,FidkowskiKitaev}, where the Majorana modes may originate from nontrivial Bogoliubov--de Gennes band structures, as Cooper pairs can be exchanged with a bulk $3$D superconductor.

{\em Symmetry analysis.}---Though our system is gapless, its topological boundary modes can be understood using the symmetry classification of $1$D \emph{gapped} fermionic phases \cite{FidkowskiKitaevClass}. We therefore begin with a discussion of the underlying symmetries. Below we argue that the Majorana boundary modes are protected by a unitary $\mz_2$ leg-interchange symmetry ${\sf L_s}$ which takes $c_{n,0}\!\rightarrow\!(-1)^{n+1}c_{n,1},\ c_{n,1}\!\rightarrow\!(-1)^{n+1}c_{n,0}$, where $n$ indexes the site along the chain. This is a symmetry only for $\phi\!=\!\pi$; we will see that its action is equivalent to that of an emergent fermion parity operator that we will describe presently using bosonization. Additionally, for any flux $\phi$, there is an antiunitary time-reversal symmetry that combines complex conjugation with interchanging the two legs of the ladder, obeying ${\sf T}^2\!=\!+1$. In principle, this enables eight distinct types of boundary modes \cite{FidkowskiKitaevClass}, though numerically we observe only one of these, with a single Majorana zero mode at each boundary. Finally, there also exists an overall $U(1)$ fermion number conservation. In our case its chief significance is that, unlike the situation considered by Ref.~\cite{FidkowskiKitaevClass}, only at exactly half-filling is a microscopic (rather than emergent) particle-hole symmetry possible. If $t_{\perp}\!=\!0$, the model has an additional exact $\mz_2$ symmetry, corresponding to the fermion parity of a single leg of the ladder, given by ${\sf P}_\ell$. It is this symmetry that protects the topological boundary modes observed by Ref.~\cite{Kraus} at zero flux.

\begin{figure}[ht]
\centering
\includegraphics[width=0.472\textwidth]{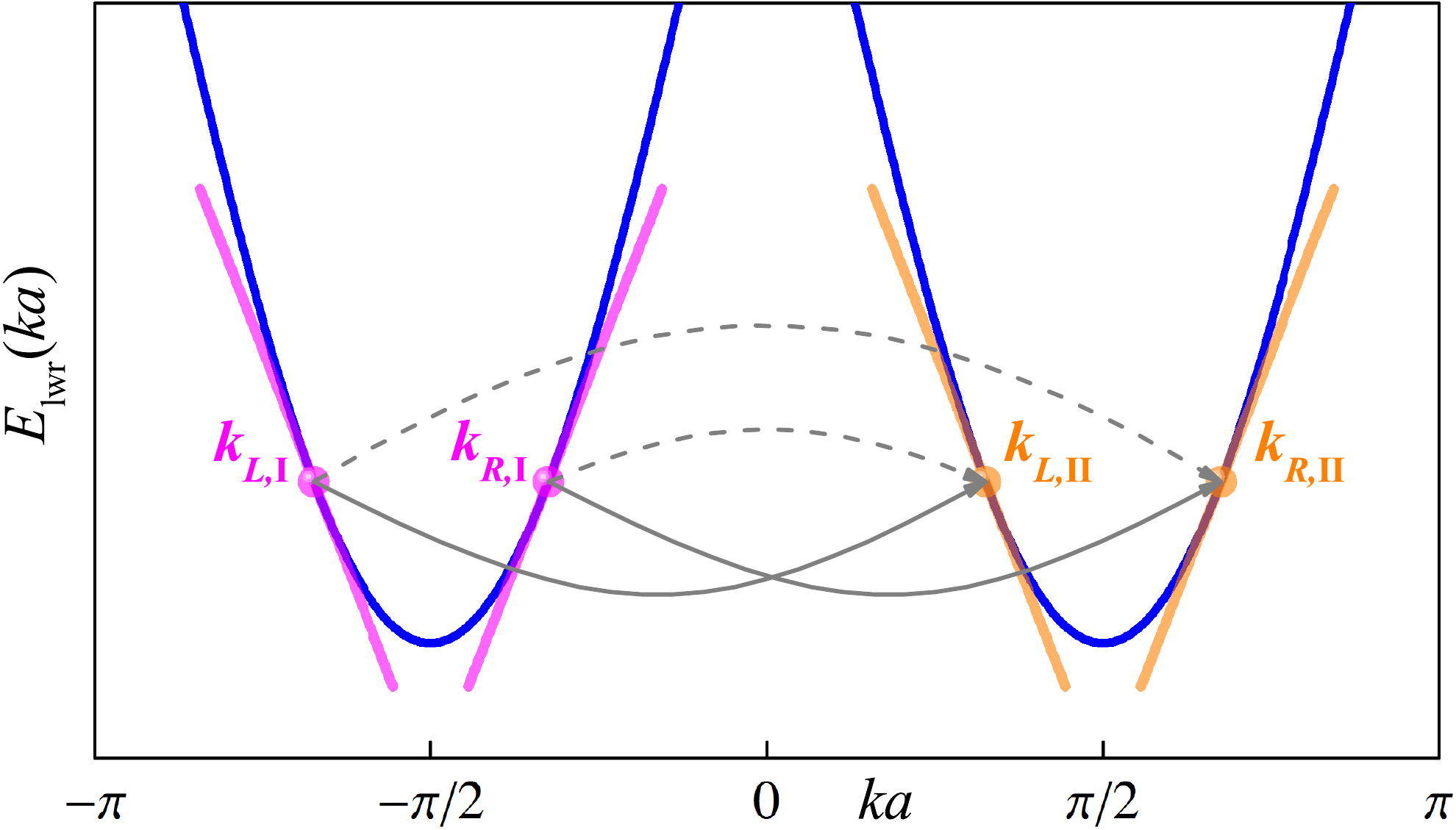}
\caption{\label{fig:fig1} Linearization of the lower band (blue solid line) at $\phi=\pi$. The obtained four chiral-fermion branches can be separated into valley-I (light magenta) and valley-II (light orange) that generalize the original chain degrees of freedom. Specifically, we depict the two types of umklapp processes that obey Eq.~(\ref{umklapp_condition}).}
\end{figure}

{\em Bosonization and renormalization group analysis.}---To understand the special role of the $\mz_2$ symmetry ${\sf L_s}$, we bosonize the model (\ref{modelhamiltonian}). For $\phi\!=\!\pi$, the kinetic Hamiltonian $H_K$ has band energies $E_{\textrm{hgr/lwr}}\!=\!\pm\!\sqrt{t^2_{\perp}\!+\!4t^2_{\parallel}\sin^2\!\left(ka\right)}$, with band gap $2t_{\perp}$. Here we consider a system at less than half-filling, with the interaction scale $W$ small relative to the bandwidth, such that we can project out the unoccupied band and focus only on the processes inside the lower band when treating $H_W$. Here the Fermi energy intersects the lower band at four separate Fermi points, as shown in Fig.~\ref{fig:fig1}.

Linearizing about these four Fermi points results in two right-moving ($R$) and two left-moving ($L$) fermion operators, which we distinguish using a valley index ($\rmI$ or $\rmII$). Bosonizing these in the usual way, we have $\psi_{\kappa,\nu}\!\sim\!e^{i\varphi_{\kappa,\nu}}$, with $\kappa\!=\!R\ (L)$ and $\nu\!=\!\rmI\ (\rmII)$. We define the nonchiral bosonic fields
\begin{align}
\theta_c&=\frac{1}{\sqrt{2}}\{\theta_{\rmI}+\theta_{\rmII}\},\ \ \theta_s=\frac{1}{\sqrt{2}}\{\theta_{\rmI}-\theta_{\rmII}\}, \\
\phi_c&=\frac{1}{\sqrt{2}}\{\phi_{\rmI}+\phi_{\rmII}\},\ \ \phi_s=\frac{1}{\sqrt{2}}\{\phi_{\rmI}-\phi_{\rmII}\},
\end{align}
where $\theta_{\nu}\!=\!\frac{1}{\sqrt{2}}(\varphi_{R,\nu}\!-\!\varphi_{L,\nu})$, $\phi_{\nu}\!=\!\frac{1}{\sqrt{2}}(\varphi_{R,\nu}\!+\!\varphi_{L,\nu})$, such that $\varphi_{\kappa,\nu}\!=\!\frac{1}{2}[\left(\phi_c\!+\!\kappa\theta_c\right)\!+\!\nu\!\left(\phi_s\!+\!\kappa\theta_s\right)]$. After including appropriate Klein factors, the only nontrivial commutators among these nonchiral bosonic fields are $\left[\theta_c(x),\phi_c(x')\right]\!=\!-2i\pi\Theta(x'\!-\!x),\ \left[\theta_s(x),\phi_s(x')\right]\!=\!2i\pi\Theta(x\!-\!x'),\ \left[\theta_s(x),\phi_c(x')\right]\!=\!-2i\pi$, where $\Theta(x)$ is the Heaviside step function. The corresponding density and current operators are: $\rho_c\!=\!(1/\pi)\partial_x\theta_c,\ \rho_s\!=\!(1/\pi)\partial_x\theta_s,\ J_c\!=\!(1/\pi)\partial_x\phi_c,\ J_s\!=\!(1/\pi)\partial_x\phi_s$.

Bosonizing the interaction term $H_W$ produces multiple four-fermion terms, of which only slowly-varying terms contribute in the continuum limit. When $\phi\!=\!\pi$, in addition to the usual momentum-conserving processes, this allows for intervalley umklapp scattering, since the Fermi points obey $k_{L,\rmI}\!=\!-k_{R,\rmII},\ k_{R,\rmI}\!=\!-k_{L,\rmII},\ k_{L,\rmII}\!+\!k_{R,\rmII}\!=\!\pi/a$, such that
\beq
k_{L,\rmII}+k_{R,\rmII}-k_{L,\rmI}-k_{R,\rmI}=\frac{2\pi}{a}.
\label{umklapp_condition}
\eeq
Eq.~(\ref{umklapp_condition}) is valid independent of the chemical potential within the lower band, but it will not hold if $\phi\neq\pi$.

The resulting bosonized form of $H$ decouples into a gapless charge sector and a gapped spin sector:
\begin{align}
H&\simeq H_c+H_s, \\
H_c&=\int_x\frac{1}{2\pi}\{u_cK_c\left(\partial_x\phi_c(x)\right)^2+\frac{u_c}{K_c}\left(\partial_x\theta_c(x)\right)^2\}, \label{H_c}\\
H_s&=\int_x\frac{1}{2\pi}\{u_sK_s\left(\partial_x\phi_s(x)\right)^2+\frac{u_s}{K_s}\left(\partial_x\theta_s(x)\right)^2\} \nonumber \\
&+\frac{2g_{um}}{(2\pi a)^2}\int_x\cos\left(2\phi_s(x)\right)-\frac{2g_{bs}}{(2\pi a)^2}\int_x\cos\left(2\theta_s(x)\right) \nonumber \\
&-\frac{2g_{mx}}{(2\pi a)^2}\int_x\cos\left(2\theta_s(x)\right)\cdot\cos\left(2\phi_s(x)\right), \label{H_s}
\end{align}
where $\int_x\!\equiv\!\int\!dx$ and the associated coupling constants are given by $g_{um}\!=\!-\frac{aW}{\pi^2}\!\cos^2(k_{R,\rmII}a)(\sin^4\!\frac{\xi}{2}\!+\!\cos^4\!\frac{\xi}{2}),\ g_{bs}\!=\!-\frac{aW}{2\pi^2}\!\sin^2(k_{R,\rmII}a)\sin^2\!\xi,\ \mbox{and}\ g_{mx}\!=\!-\frac{aW}{2\pi^2}(\sin^4\!\frac{\xi}{2}\!+\!\cos^4\!\frac{\xi}{2})$. Here the wavefunction in the lower band at $k_{R,\rmII}$, has the form $(\cos\!\frac{\xi}{2},\ \sin\!\frac{\xi}{2})$ in the leg (i.e., $0,1$) basis \cite{SupplementalMaterial}. Note that the $g_{um}$-term favors Cooper pairing while the competing $g_{bs}$-term favors a valley density-wave-type order. Moreover, owing to the density-density interactions, the velocities and Luttinger parameters are renormalized. Particularly, for the spin channel
\begin{align}
u_s&\!=\!\sqrt{(u\!+\!g)^2\!-\!4g^2\cos^2(2k_{R,\rmII}a)}, \\
K_s&\!=\!\sqrt{\frac{u\!+\!g\!+\!2g\cos(2k_{R,\rmII}a)}{u\!+\!g\!-\!2g\cos(2k_{R,\rmII}a)}},
\end{align}
where $u\!=\!\frac{1}{2}v_F\!\equiv\!\frac{1}{2}\frac{dE_{\textrm{lwr}}}{dk}\big|_{k=k_{R,\rmII}}\!>\!0$ and $g\!=\!\frac{aW}{2(2\pi)^3}\!\sin^2\!\xi$.

As the charge sector is gapless, we will concentrate on the spin sector, which is responsible for the topological Majorana boundary modes. To determine which of the sine-Gordon terms in $H_s$ dominates, we use the one-loop renormalization group (RG) flow equations:
\begin{align}
\frac{dK_s(l)}{dl}&=\frac{y^2_{um}(l)}{2}-\frac{y^2_{bs}(l)K^2_s(l)}{2}, \\
\frac{dy_{um}(l)}{dl}&=(2-2K^{-1}_s(l))y_{um}(l), \\
\frac{dy_{bs}(l)}{dl}&=(2-2K_s(l))y_{bs}(l), \\
\frac{dy_{mx}(l)}{dl}&=(2-2K_s(l)-2K^{-1}_s(l))y_{mx}(l),
\end{align}
where $y_{um}\!=\!\frac{g_{um}}{2u\pi},\ y_{bs}\!=\!\frac{g_{bs}}{2u\pi},$ and $y_{mx}\!=\!\frac{g_{mx}}{2u\pi}$ are the dimensionless coupling constants. The $g_{mx}$-channel, involving both $\theta_s$ and $\phi_s$, is power-counting irrelevant for any value of $K_s$ and can be neglected. Accordingly, the gap-opening competition is between $g_{um}$ and $g_{bs}$. If we take $W$ to be negative, then for $3\pi/(4a)\!>\!k_{R,\rmII}\!>\!\pi/(2a)$ we have $K_s\!>\!1$ and $g_{um}$ is the only relevant coupling. In this regime the long-wavelength effective Hamiltonian is therefore expected to be
\begin{align}
H_{s}\!\sim\!H_{\textrm{s-G}}&\!=\!\!\int^{\frac{L}{2}}_{-\frac{L}{2}}\frac{dx}{2\pi}\left\{u_sK_s\left(\partial_x\phi_s(x)\right)^2+\frac{u_s}{K_s}\left(\partial_x\theta_s(x)\right)^2 \right. \nonumber \\
&\left. \ \ \ \ \ \ \ \ \ \ \ \ \ \ \ \ \ \ \ \ \ \ \ \ +\frac{g_{um}}{\pi a^2}\cos\left(2\phi_s(x)\right)\right\}.
\label{H_s-G_pi}
\end{align}
Note that for $W\!<\!0,\ g_{um}\!>\!0$ and the coefficient of the $\cos\!\left(2\phi_s\right)$ term is positive, contrary to the usual sine-Gordon model. As explained below, this sign flip, which arises from the particular umklapp processes that occur in the band structure for $\phi\!=\!\pi$, is crucial to ensuring that the staggered leg-interchange symmetry ${\sf L_s}$ acts like a fermion parity operator. Indeed, for $W\!>\!0$ and $3\pi/(4a)\!>\!k_{R, \rmII}\!>\!\pi/(2a)$ it is the negative coupling $g_{bs}$ that is most relevant; this phase shows no numerical signatures of topological boundary modes, as discussed in the Supplemental Material \cite{SupplementalMaterial}.

{\em $\mz_2$ symmetry and Majorana operators.}---Following Ref.~\cite{Clarke} we can further derive the bosonized Majorana-boundary-mode operators by specifying the vacuum Hamiltonian density ${\sf H_{vac}}(x)\!=\!-\frac{2M_{\infty}}{\pi a}\!\sin\!\left(\theta_s(x)\right)\!\cos\!\left(\theta_c(x)\right)$, which amounts to sending the fermion mass outside the system to $+\infty$ on both legs of the ladder \cite{Keselman}. This fixes $\theta_s(x\!\in\!(-\infty,-\frac{L}{2}))\!=\!\pm\frac{\pi}{2}\!+\!2\hat{n}^{(1)}_{\theta_s}\pi$ and $\theta_s(x\!\in\!(\frac{L}{2},\infty))\!=\!\pm\frac{\pi}{2}\!+\!2\hat{n}^{(2)}_{\theta_s}\pi$. In the gapped bulk of the ladder, $\phi_s(x\!\in\!(-\frac{L}{2},\frac{L}{2}))\!=\!-\frac{\pi}{2}\!+\!\hat{n}_{\phi_s}\pi$, where $\hat{n}^{(1,2)}_{\theta_s},\ \hat{n}_{\phi_s}$ are integer operators. The two domain-wall Majorana operators are then: $\gamma_L\!\simeq\!e^{i\pi(\hat{n}^{(1)}_{\theta_s}+\hat{n}_{\phi_s})},\ \gamma_R\!\simeq\!e^{i\pi(\hat{n}^{(2)}_{\theta_s}+\hat{n}_{\phi_s})}$. These satisfy the Majorana relations $\gamma^{\dagger}_{L/R}\!=\!\gamma_{L/R},\ \gamma^2_{L/R}\!=\!1,\ \{\gamma_L,\gamma_R\}\!=\!0$, and define an emergent fermion parity $\Pv\!=\!i\gamma_L\gamma_R$. This acts on the spin fields within the gapped system via $\Pv\phi_s{\sf P^{-1}_v}\!=\!\phi_s\!-\!\pi,\ \Pv\theta_s{\sf P^{-1}_v}\!=\!\theta_s$. In particular, when $g_{um}$ dominates the RG flow, $\Pv$ maps between the system's two classical ground states; when $g_{bs}$ dominates the flow, it acts trivially on the two classical ground states.

Strictly speaking the operator $\Pv$ does not correspond to any microscopic symmetry for $|t_\perp|\!>\!0$. (For $t_{\perp}\!=\!0$, it can be interpreted as the fermion parity of one of the ladder's two legs.) However, at $\phi\!=\!\pi$ its action is equivalent to that of the microscopic $\mz_2$ symmetry $\Ls$, which acts on the bosonized field via $\Ls\phi_{s}(x){\sf L^{-1}_s}\!=\!-\phi_{s}(x)\ (\mbox{mod}\ 2\pi)$. Specifically, since $g_{um}$ is positive, both $\Pv$ and $\Ls$ map between the two classical minima $\left|\pm\pi/2\right\rangle$ of the sine-Gordon potential, where $\phi_s(x)\left|\pm\pi/2\right\rangle\!=\!\left(\pm\pi/2\ \mbox{mod}\ 2\pi\right)\left|\pm\pi/2\right\rangle$. Thus within the ground-state manifold, the action of $\Pv$ is equivalent to that of the exact symmetry of $\Ls$. Indeed, in a finite-size system, due to the proliferation of instanton processes \cite{ChenBurnell,Fidkowski}, the ground states of $H_{\textrm{s-G}}$ are split into $\left|\pm\right\rangle\!=\!\frac{1}{\sqrt{2}}\!\left(\left|\pi/2\right\rangle\!\pm\!\left|-\pi/2\right\rangle\right)$ whose energy difference is exponentially small in the system's size. The lowest two eigenstates therefore obey $\Pv\left|\pm\right\rangle\!=\!\pm\left|\pm\right\rangle,\ \Ls\left|\pm\right\rangle\!=\!\pm\left|\pm\right\rangle$, and can be labeled by their eigenvalues under the microscopic symmetry $\Ls$.

We note that for flux $\phi\!\neq\!\pi$, the umklapp scattering leading to Eq.~(\ref{H_s-G_pi}) is no longer an effective zero (lattice)-momentum transfer process, and $\cos\!\left(2\phi_s(x)\right)$ is replaced by $\cos\!\left(2\phi_s(x)\!-\!\delta x\right)$, where $\delta$ parameterizes the deviation of the flux from $\pi$. In this case we expect a commensurate-incommensurate transition which destroys the topological boundary modes. As described in the Supplemental Material \cite{SupplementalMaterial}, numerical evidence suggests that this transition leads to a phase in which $\theta_s$ is locked.

{\em Comparing $\phi\!=\!\pi$ with $\phi\!=\!0$.}---As noted above, the topological state found in Ref.~\cite{Kraus} at $\phi\!=\!0,\ t_{\perp}\!=\!0$ is protected by the symmetry ${\sf P_\ell}$ corresponding to the conserved fermion parity of one of the ladder's legs. For $t_\perp\!=\!0$, the operator $\Pv$ constructed from the Majorana boundary modes is exactly equal to ${\sf P_\ell}$, and therefore corresponds to an exact symmetry that protects the resulting topological boundary modes. In comparison, at $\phi\!=\!\pi$ we have found that $\Pv$'s action on the classical ground states is identical to that of the leg-exchange symmetry $\Ls$, which is not violated by finite $t_{\perp}$. Hence it is the particular action of $\Ls$ at this point which gives the topological boundary modes their enhanced stability. This result highlights the fact that different choices of topologically trivial band structures can have profound implications for interaction-driven topological phases.

In addition to these symmetry considerations, introducing $t_{\perp}$ at $\phi=0$ has a significantly different impact on the band structure and umklapp processes than doing so at $\phi\!=\!\pi$, making the latter state more stable. For $\phi\!=\!t_{\perp}\!=\!0$, the two chains' bands are identical. This overlap favors the umklapp-scattering processes that lead to the topological phase. However, increasing $t_{\perp}$ at $\phi\!=\!0$ separates the two bands and creates a Fermi-velocity mismatch, which disfavors these processes, ultimately causing the spin gap to close at a small but finite value of $t_\perp$ \cite{Kraus}. By contrast, when $\phi\!=\!\pi$, the two valleys where the chains' bands cross the Fermi surface are maximally separated by a wave vector $\pi/a$, and remain symmetric with a finite $t_{\perp}$. In this case the Fermi-velocity mismatch is absent, and the spin gap persists up to large values of $t_{\perp}$. Thus the different nontopological band structures play a key role in making the topological Majorana modes robust (fragile) against $t_{\perp}$ at $\phi\!=\!\pi$ ($\phi\!=\!0$), consistent with the fact that $t_{\perp}$ breaks ${\sf P_\ell}$ but preserves ${\sf L_s}$.

\begin{figure}[ht]
\centering
\includegraphics[width=0.475\textwidth]{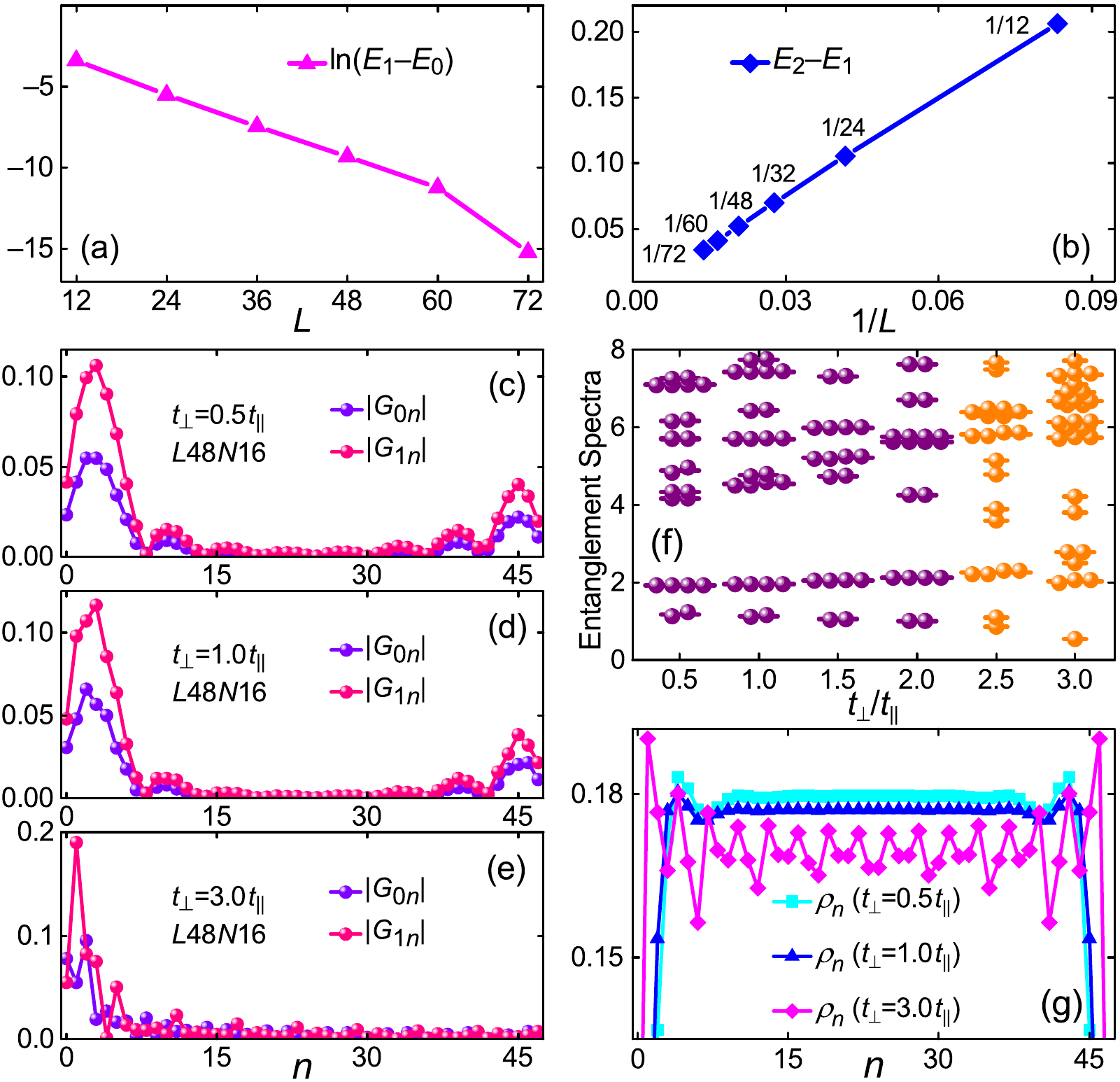}
\caption{\label{fig:fig2} Numerical signatures of the topological phase. (a) and (b): Scaling of energy gaps as functions of $L$ from DMRG. (a) shows that the energy difference between the first two lowest-lying eigenstates of (\ref{modelhamiltonian}) decays exponentially with $L$. The protected ground-state manifold is separated from the rest of the spectrum by a gap that decreases inversely with $L$, as shown in (b). Here $W\!=\!-1.7t_{\parallel},\ t_{\perp}\!=\!0.5t_{\parallel},\ \phi\!=\!\pi,\ N/L\!=\!\!1/3$. (c)--(g): Transition out of the topological phase at large $t_{\perp}$ for fixed $W\!=\!-1.7t_{\parallel},\ \phi\!=\!\pi,\ L\!=\!48,\ N\!=\!16$. (c)--(e) demonstrate the edge mode via the nonlocal correlations \cite{Kraus,Lang,Iemini} in single-particle Green functions. At $t_{\perp}\!=\!3.0t_{\parallel}$, the edge mode disappears indicating the transition to a trivial state. This is in accordance with (f) and (g) which show the corresponding evolutions of entanglement spectra and local fermion densities as the transition is approached.}
\end{figure}

{\em Numerical verification.}---Simulations based on density matrix renormalization group (DMRG) \cite{White} and exact diagonalization (ED) \cite{Sandvik} have been performed to solve the lattice model (\ref{modelhamiltonian}) at $\phi\!=\!\pi$ and $t_{\perp}\!=\!0.5t_{\parallel}$. The numerical outcomes provide strong evidence supporting our theoretical predictions. Fig.~\ref{fig:fig2}(a) demonstrates that in the low-population region ($N/L\!=\!1/3$), when pair hopping is strong ($W\!=\!-1.7t_{\parallel}$), the energy gap between the ground state and the $1$st excited state closes exponentially as the ladder's size $L$ increases. This is in contrast with a power-law gap closing resulting from the gapless charge sector. As anticipated, these two nearly degenerate eigenstates are distinguished by their eigenvalues of $\Ls$: The ground ($1$st excited) state has eigenvalue $+1$ ($-1$). The resulting ground-state manifold is further separated from the rest of the spectrum by a gap which only decreases inversely with $L$ (Fig.~\ref{fig:fig2}(b)). Moreover, the topological Majorana boundary modes can be characterized via the nonlocal correlations \cite{Kraus,Lang,Iemini} in the single-particle Green functions $G_{mn}\!\coloneqq\!\langle c^{\dagger}_{m,0}c_{n,0}\rangle\!=\!\langle c^{\dagger}_{m,1}c_{n,1}\rangle$. These are apparent for a range of $t_\perp$ values in the topological regime (Figs.~\ref{fig:fig2}(c)--(d)). The presence of the edge states also gives rise to a two-fold degeneracy in the entanglement spectrum (ES) \cite{Li,Pollmann} on the central bond (Fig.~\ref{fig:fig2}(f)). These DMRG results have been confirmed by ED for small system sizes.

As discussed in the Supplemental Material \cite{SupplementalMaterial}, adding a small $\Ls$-symmetry-breaking perturbation to (\ref{modelhamiltonian}) causes the degeneracy in the ES to lift, and the energy gap of $E_1\!-\!E_0$ to deviate from the exponentially small splitting observed in Fig.~\ref{fig:fig2}(a). This supports our claim that the topological boundary modes are protected by $\Ls$ at $\phi\!=\!\pi$ and $t_{\perp}\!\neq\!0$.

To fully characterize the topological regime, we also show evidence of a phase transition out of the topological regime at $t_{\perp,\textrm{c}}\!\approx\!2.5t_{\parallel}$. Fig.~\ref{fig:fig2}(e) shows a value $t_{\perp}\!>\!t_{\perp,\textrm{c}}$, for which the nonlocal correlations in $G_{mn}$ are absent. Further, as shown by Fig.~\ref{fig:fig2}(f), the degeneracy in the ES apparent for $t_{\perp}\!<\!t_{\perp,\textrm{c}}$ is lifted and the lowest level is clearly nondegenerate at $t_{\perp}\!=\!3.0t_{\parallel}$. Our numerics suggest that the transition is toward a state with density-wave order at large $t_{\perp}$: For $t_{\perp}\!>\!t_{\perp,\textrm{c}}$, the fermion-density profile $\rho_n\!\coloneqq\!\langle c^{\dagger}_{n,0}c_{n,0}\rangle\!=\!\langle c^{\dagger}_{n,1}c_{n,1}\rangle$ in Fig.~\ref{fig:fig2}(g) evolves from a uniform distribution toward an oscillatory pattern. Additionally, for $t_{\perp}\!\geq\!3.0t_{\parallel}$, the two lowest energy eigenstates split and both have $\Ls$ eigenvalues of $+1$ \cite{SupplementalMaterial}. All the above observations suggest that once $t_{\perp}\!>\!t_{\perp,\textrm{c}}$, the Majorana boundary modes disappear.

{\em Experimental feasibility.}---In cold atom laboratories, currently a great deal of effort has been devoted to simulating strong synthetic magnetic fields in optical lattices by Raman-assisted tunnelings and lattice-shaking techniques, which mimic standard Peierls substitutions through attaching an Aharanov--Bohm-like complex phase to the nearest-neighbor single-particle hopping \cite{Aidelsburger,AidelsburgerFlux,MiyakeFlux,Atala,Mancini,Kennedy,Struck,Livi}. The remarkable tunability of these artificial gauge fluxes has paved an experimental avenue to exploring the Bose--Einstein condensation in Harper--Hofstadter model of $2$D bosonic gases \cite{Kennedy}. Chiral edge states due to particle's cyclotron motion stirred by uniform fluxes have also been detected in quasi-$1$D fermionic ladders and Hall ribbons, where a sizable flux $\phi$ can reach up to $1.31\pi$ per plaquette by utilizing the technology of optical atomic clocks \cite{Livi,Mancini}. In Ref.~\cite{Kraus} an atomic scheme has also been outlined for creating the more intricate pair-hopping interaction. These developments make the prospect of realizing flux-stabilized Majorana zero modes a possibility in the near future.

To summarize, we have shown that by threading $\pi$-flux through each plaquette, interaction-driven Majorana bound states in fermionic ladders may be stabilized in the presence of single-particle interleg tunneling. \emph{En~route} we have established a connection between a microscopic $\mz_2$ leg-exchange symmetry present only at this flux value and the action of an emergent fermion parity operator in the long-wavelength bosonized theory. We have also highlighted the advantages of the $\pi$-flux state in fostering umklapp processes which generate the spin gap enabling this topological regime. Our theory has been substantiated by extensive DMRG and ED calculations.

C.~C. thanks M.~D.~Schulz for discussions. Part of the simulation is developed from ALPS package \cite{Dolfi}. C.~C. and F.~J.~B. are supported by NSF-DMR $1352271$ and by the Sloan Foundation FG-$2015$-$65927$. W.~Y. and Y.~C. are supported by the National Natural Science Foundation of China (Grant Nos. $11625416$, $11474064$, and $11274069$). C.~S.~T. is supported by the Robert A. Welch Foundation under Grant No.~E-$1146$.

C.~C. and W.~Y. contributed equally to this work.

\onecolumngrid

\appendix

\vspace{6mm}

\begin{center}
\large{\textbf{{Supplemental Material}}}
\end{center}

\vspace{-6mm}

\section{Band structure}

Here we provide more details on the band structure of $H_K$, and show how it enters the bosonized form of the full Hamiltonian. The noninteracting kinetic Hamiltonian,
\begin{align}
H_K&=\int^{\infty}_{-\infty}\frac{dk}{2\pi}\left(\psi^{\dagger}_{k,0}\ \ \ \psi^{\dagger}_{k,1}\right)\left(
\begin{array}{cc}
 -2t_\parallel\cos\!\left(ka+\frac{\phi}{2}\right) & -t_{\perp} \\
 -t_{\perp} & -2t_\parallel\cos\!\left(ka-\frac{\phi}{2}\right) \\
\end{array}
\right)\left(
\begin{array}{c}
 \psi_{k,0} \\
 \psi_{k,1} \\
\end{array}
\right) \nonumber \\
&=\int^{\infty}_{-\infty}dk\left(\psi^{\dagger}_{k,+}\ \ \ \psi^{\dagger}_{k,-}\right)\left(
\begin{array}{cc}
 E_+(k,\phi) & 0 \\
 0 & E_-(k,\phi) \\
\end{array}
\right)\left(
\begin{array}{c}
 \psi_{k,+} \\
 \psi_{k,-} \\
\end{array}
\right), \nonumber
\end{align}
can be diagonalized by introducing the unitary matrix $\mathbb{M}(k,\phi)$,
\beq
\mathbb{M}(k,\phi)=\left(
\begin{array}{cc}
 |\chi_+\rangle & |\chi_-\rangle \\
\end{array}
\right)=\left(
\begin{array}{cc}
 \frac{-u_{k,\phi}}{\sqrt{1+u_{k,\phi}^2}} & \frac{1}{\sqrt{1+u_{k,\phi}^2}} \\ [0.2em]
 \frac{1}{\sqrt{1+u_{k,\phi}^2}} & \frac{u_{k,\phi}}{\sqrt{1+u_{k,\phi}^2}} \\
\end{array}
\right),\ \ \ \mathbb{M}^{\dagger}\mathbb{H}(k,\phi)\mathbb{M}=\left(
\begin{array}{cc}
 E_+(k,\phi) & 0 \\
 0 & E_-(k,\phi) \\
\end{array}
\right), \nonumber
\eeq
where the component in the matrix elements assumes
\beq
u_{k,\phi}=\frac{1}{t_{\perp}}\left\{2t_{\parallel}\sin{(ka)}\sin\frac{\phi}{2}+\sqrt{t^2_{\perp}+\left(2t_{\parallel}\sin(ka)\sin\frac{\phi}{2}\right)^2}\right\}. \nonumber
\eeq

From now on, we focus on the case of $\phi=\pi$, with the chemical potential in the lower band. Linearizing the band structure about the four Fermi points in the lower band gives rise to four independent but mutually interacting chiral-fermion branches, which we label $(L,\rmI),\ (R,\rmI),\ (L,\rmII),$ and $(R,\rmII)$ (see Fig.~1 in the main text). The fermion operator in real space is related to these four chiral fermions via:
\begin{align}
\psi_{x,\ell}&\simeq e^{\imi k_{L,\rmI}x}\langle \ell|\chi_-(k_{L,\rmI})\rangle\tpsi_{L,\rmI}(x)+e^{\imi k_{R,\rmI}x}\langle \ell|\chi_-(k_{R,\rmI})\rangle\tpsi_{R,\rmI}(x) \nonumber \\
&+e^{\imi k_{L,\rmII}x}\langle \ell|\chi_-(k_{L,\rmII})\rangle\tpsi_{L,\rmII}(x)+e^{\imi k_{R,\rmII}x}\langle \ell|\chi_-(k_{R,\rmII})\rangle\tpsi_{R,\rmII}(x), \nonumber
\end{align}
where $x$ is the position along the ladder, and $\ell=0,1$ indexes the leg. Here we have defined $\langle\ell=0|\equiv\left(1\ \ 0\right)$, $\langle \ell=1|\equiv\left(0\ \ 1\right)$, and $|\chi_-(k)\rangle\equiv|\chi_-(k,\phi=\pi)\rangle=\left(
\begin{array}{c}
 \frac{1}{\sqrt{1+u^2_{k,\pi}}} \\ [0.2em]
 \frac{u_{k,\pi}}{\sqrt{1+u^2_{k,\pi}}} \\
\end{array}
\right)=\left(
\begin{array}{c}
 \cos\frac{\xi(k)}{2} \\ [0.3em]
 \sin\frac{\xi(k)}{2} \\
\end{array}
\right)$.

The Hamiltonian in the main text is obtained by substituting these expressions into the pair-hopping term $H_W$, and performing a standard bosonization analysis of the result. Following these calculations through, one finds that the matrix elements $\langle \ell | \chi_- \rangle$ enter the resulting Luttinger parameters, as well as the sine-Gordon couplings $g_{um},\ g_{bs},\ g_{mx}$. The relevant expressions in the spin sector are given in the main text; in the charge sector the associated Luttinger parameters are defined by
\begin{align}
u_c&=\sqrt{(u-g)\cdot(u-5g)},\ \ \ \ \ K_c=\sqrt{\frac{u-g}{u-5g}}, \nonumber
\end{align}
where the matrix element enters $u_c$ and $K_c$ via $g\!=\!\frac{aW}{2(2\pi)^3}\sin^2\xi$.

It follows that the microscopic parameters in our bosonized model are sensitive to the precise initial band structure, as well as to the initial value of the chemical potential within the lower band. In particular, the wrong choice of chemical potential can lead to $K_s<1$, and an RG flow that tends to take the system away from the topological regime.

\section{Topological quantum phase transition}

Though at $\phi=\pi$ our model is robust to the single-particle interchain tunneling up to $t_{\perp}\approx t_{\parallel}$, Figs.~2(c)--(e) in the main text suggest that by $t_{\perp}=3t_{\parallel}$, the system has undergone a transition to a nontopological regime. In particular, for this value we see that the nonlocal correlation in the single-particle Green function, an indicator of the presence of Majorana boundary modes, disappears. Further, the double degeneracy in the entanglement spectrum, another indicator of the topological boundary modes, disappears by $t_{\perp}=2.5t_{\parallel}$. Finally, the spatial profile of the local fermion density (shown in Fig.~2(g) in the main text) shows clear evidence of a fermion density-wave-type order for $t_{\perp}=3t_{\parallel}$. In comparison, the bulk fermion density is uniform for $t_{\perp}\leq t_\parallel$, where the nonlocal correlation indicates the presence of Majorana boundary modes.

\begin{figure}[ht]
\centering
\includegraphics[width=0.6\textwidth]{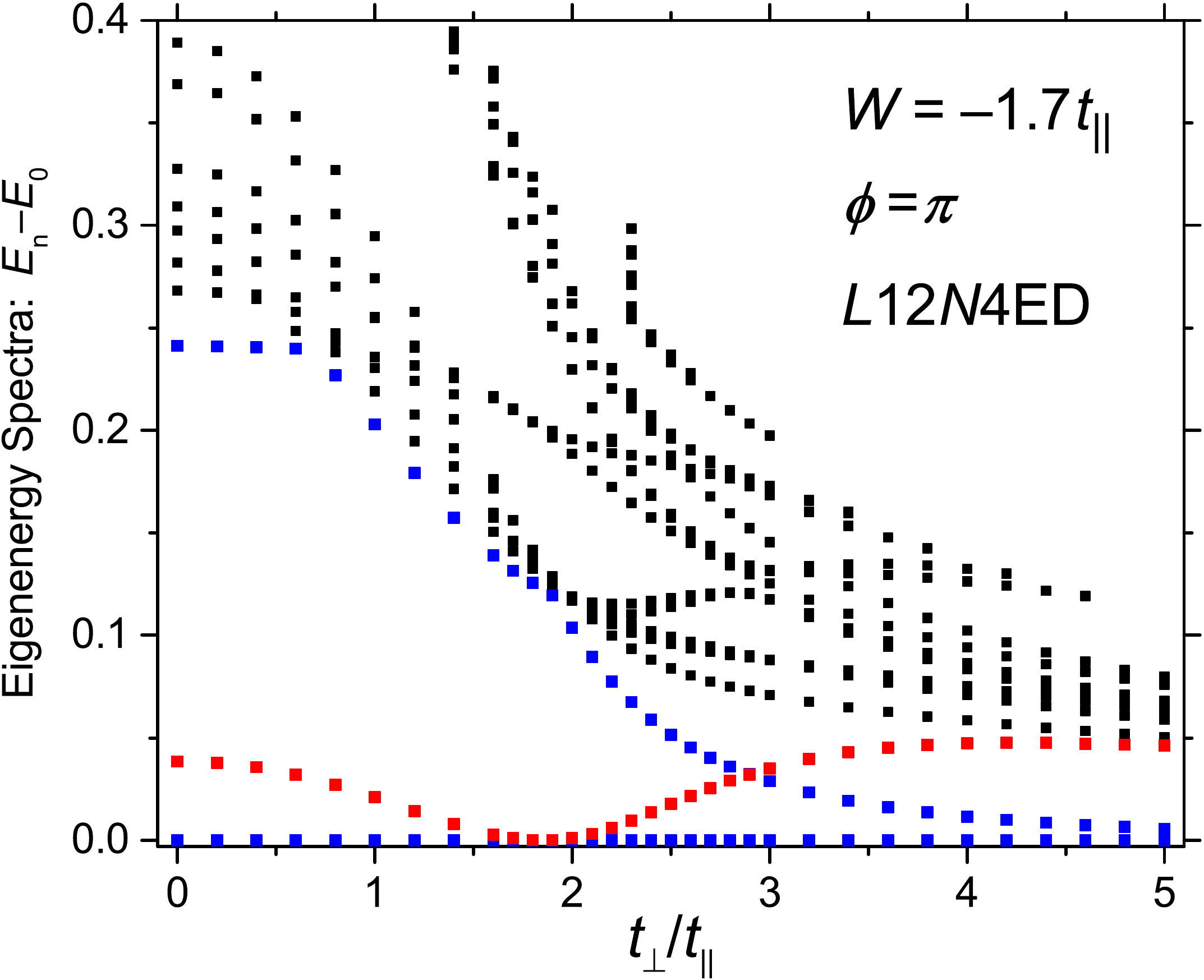}
\caption{\label{fig:fig1_supp} Exact diagonalization results show the evolution of the rescaled eigenenergies of $H$: $E_n-E_0$, as we increase $t_{\perp}$. Here we fixed $W=-1.7t_{\parallel},\ \phi=\pi,\ L=12,\ N=4$, and set $t_{\parallel}=1.0$ as the energy unit. We monitor the ${\sf L_s}$-quantum number for the lowest three eigenstates and use different colors to distinguish the $\mathbb{Z}_2$ eigenvalues of ${\sf L_s}$: Blue denotes $\left\langle{\sf L_s}\right\rangle=+1$, while red denotes $\left\langle{\sf L_s}\right\rangle=-1$. When $0\leq t_{\perp}/t_{\parallel}\lesssim2.5$, the lowest two eigenstates can be differentiated by their quantum numbers of ${\sf L_s}$. However, when $t_{\perp}/t_{\parallel}\gtrsim3$, they possess the same quantum number $\left\langle{\sf L_s}\right\rangle=+1$ as the consequence of level crossing between the $1$st and $2$nd excited states.}
\end{figure}

As further evidence of this transition, Fig.~\ref{fig:fig1_supp} shows the low-energy spectrum of the $\pi$-flux ladder as a function of $t_{\perp}/t_{\parallel}$ at fixed pair-hopping strength $W$. At approximately $t_{\perp}=2.8 t_{\parallel}$ a level crossing occurs between two states with opposite $\mathbb{Z}_2$ quantum numbers under the staggered leg-interchange symmetry ${\sf L_s}$. For smaller $t_{\perp}$ the two lowest energy states have opposite $\mathbb{Z}_2$ quantum numbers; whereas for larger $t_{\perp}$ both have the same ($+1$) $\mathbb{Z}_2$ eigenvalue. As explained in the main text this $\mathbb{Z}_2$ quantum number is associated with the fermion parity in the topological region, suggesting that this level crossing indicates a transition out of the topological regime. This level crossing is consistent with what we expect for a transition from a topological regime to a fermion-density-wave order: In the latter case we expect the lowest two states to be related by translation, and hence to have the same parity under the leg-exchange symmetry ${\sf L_s}$.

\section{The pi-flux ladder for positive $W$}

The RG analysis in the main text suggests that for finite $t_{\perp}$ when $W>0,\ \phi=\pi$, the most relevant gapping channel of $H_s$ for low fermion density $\left(3\pi/(4a)>k_{R,\rmII}>\pi/(2a)\right)$ is the backscattering term $g_{bs}$. In this section we present the corresponding DMRG results to illustrate that the resulting phase exhibits no signatures of topological protection at finite $t_{\perp}$.

\begin{figure}[ht]
\centering
\includegraphics[width=0.55\textwidth]{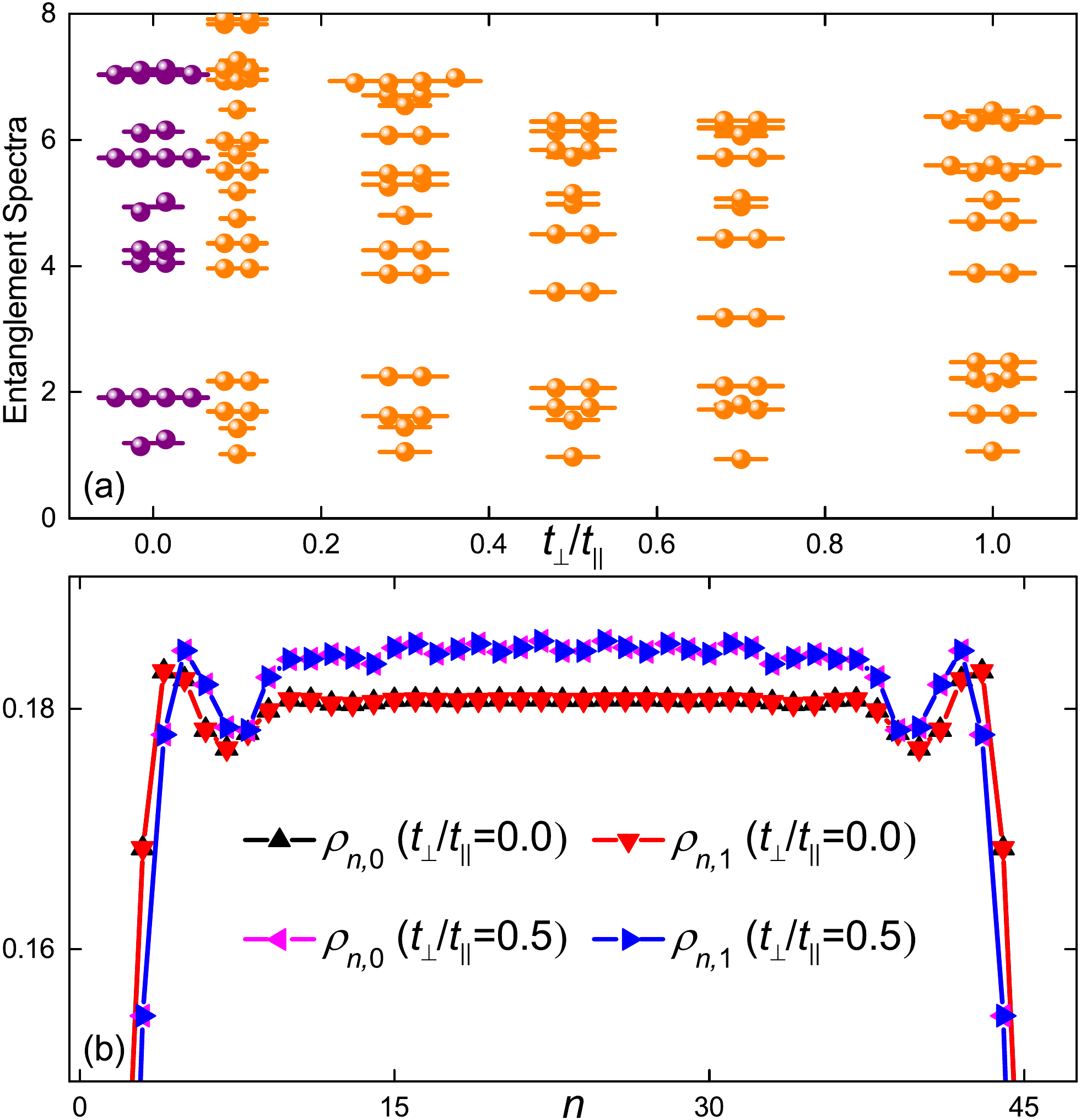}
\caption{\label{fig:fig_supp_W_pos} DMRG results: (a) shows the degeneracy splittings in the entanglement spectra as a function of $t_{\perp}$ for the ground state of the ladder system with the fixed parameters: $W=+1.7t_{\parallel},\ \phi=\pi,\ L=48,\ N=16$. Two prototypical fermion-density profiles have been illustrated in panel (b), where for the trivial state there exist weak density modulations along the chain.}
\end{figure}

In particular, we evaluate the entanglement spectra (ES) for $W=+1.7t_{\parallel}$ at several different values of $t_{\perp}$. The results in Fig.~\ref{fig:fig_supp_W_pos}(a) demonstrate the significant splitting of the ES two-fold degeneracy at the lowest level even for the smallest nonzero values of $t_{\perp}\approx0.1t_{\parallel}$. This is in sharp contrast to the situation of $W=-1.7t_{\parallel}$, where the even degeneracy in ES persists up to $t_{\perp}\approx2.0t_{\parallel}$ (see Fig.~2(f) in the main text). Further, for parameters where this degeneracy is lifted, the local fermion-density profile $\rho_{n,\ell}$ exhibits weak modulations along the chain, as can be seen from Fig.~\ref{fig:fig_supp_W_pos}(b). Here we define $\rho_{n,0}=\langle c^{\dagger}_{n,0}c_{n,0}\rangle,\ \rho_{n,1}=\langle c^{\dagger}_{n,1}c_{n,1}\rangle$ for the upper and lower chains, respectively.

\section{Leg-interchange-symmetry-breaking perturbations}

Here, we demonstrate numerically that the $\mathbb{Z}_2$ leg-exchange symmetry and the flux $\phi=\pi$ are both central to protecting the topological features of our model. Specifically, we investigate three different ways of breaking the leg-interchange symmetry, and show that small (but finite) values of all symmetry-breaking perturbations lift the degeneracies in the ES associated with the topological boundary modes. In addition, we perform an analysis of the effect of perturbing the flux away from $\phi=\pi$, which lifts the degeneracies in the entanglement spectrum and leads to a fermion-density-wave order.

\subsection{Making the intrachain hoppings different}

\begin{figure}[ht]
\centering
\includegraphics[width=0.75\textwidth]{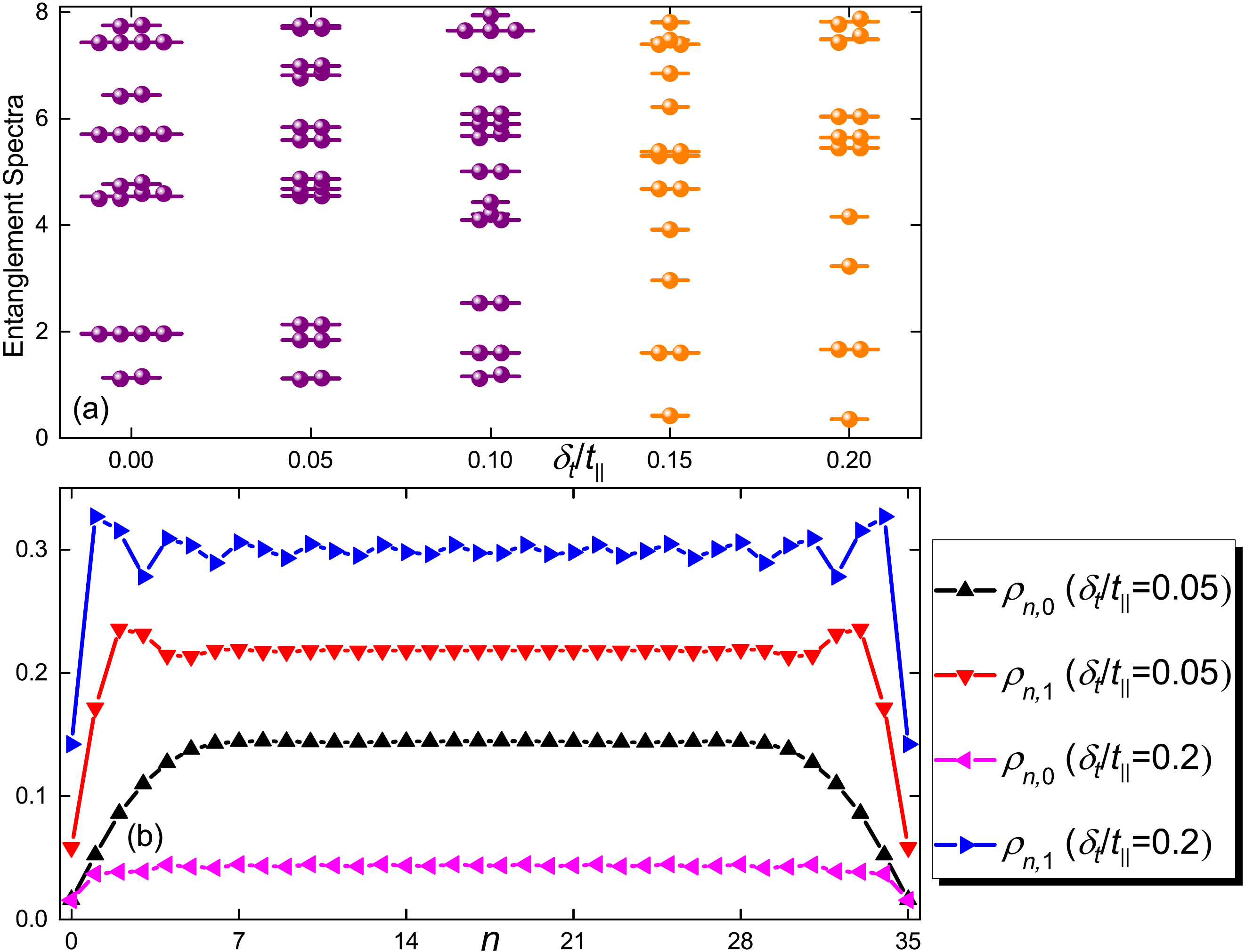}
\caption{\label{fig:fig_supp_diff_t_parallel_den} DMRG results: (a) shows the evolution of ES (for a system of $L=36,\ N=12$) as a function of ${\sf L_s}$-breaking perturbation $\delta_t/t_{\parallel}$ (see Eq.~(\ref{Hamdifftparallel})). The even degeneracy of ES survives only up to small values of $\delta_t/t_{\parallel}\approx0.125$. Panel (b) illustrates the local fermion-density profiles for the two different values of $\delta_t/t_{\parallel}$. Here $W=-1.7t_{\parallel},\ t_{\perp}=t_{\parallel},\ \phi=\pi$.}
\end{figure}

We first consider the result of explicitly breaking the leg-interchange symmetry by choosing $t_{\parallel}$ to be different on the two legs of the ladder. We thus modify the Hamiltonian in the following way:
\begin{align}
H'=&-\sum^{L-2}_{n=0}\left[\left(i\left(t_{\parallel}-\delta_t\right)  c^{\dagger}_{n,0}c_{n+1,0}-i\left(t_{\parallel}+\delta_t\right) c^{\dagger}_{n,1}c_{n+1,1}\right)+\textrm{H.c.}\right] -\sum^{L-1}_{n=0}\left(t_{\perp}c^{\dagger}_{n,0}c_{n,1} +\textrm{H.c.}\right) \nonumber \\
&+\sum^{L-2}_{n=0}\left(Wc^{\dagger}_{n,0}c^{\dagger}_{n+1,0} c_{n,1}c_{n+1,1}+\textrm{H.c.}\right),
\label{Hamdifftparallel}
\end{align}
where we have set $\phi=\pi$, and the added $\delta_t$-terms explicitly break the ${\sf L_s}$ symmetry.

The arguments given in the main text assert that breaking ${\sf L_s}$ should lift the degeneracy associated with the topological boundary modes, since there is no longer any reason for the fermion parity even and fermion parity odd states to be proximate in energy. This is apparent in the evolution of the ground-state entanglement spectra (shown in Fig.~\ref{fig:fig_supp_diff_t_parallel_den}(a) for a system of $L=36,\ N=12$) as a function of $\delta_t/t_{\parallel}$. We observe that for fixed $W=-1.7t_{\parallel},\ t_{\perp}=t_{\parallel}$, the two-fold degeneracy in the lowest level of the ES has completely disappeared by $\delta_t=0.15t_{\parallel}$. This suggests that the symmetry ${\sf L_s}$ is, as claimed, integral to protecting the topological boundary modes. Fig.~\ref{fig:fig_supp_diff_t_parallel_den}(b) further supports this, by showing that for sufficiently large $\delta_t$ ($\delta_t=0.2t_{\parallel}$), the fermion density on the lower chain tends to develop weak spatial modulations, similar to those observed for positive $W$ in the previous section. For smaller values ($\delta_t=0.05t_{\parallel}$), the density profiles appear to remain uniform in the bulk, though the relative populations of the two chains are generically different when the leg-interchange symmetry is broken.

It is somewhat surprising that for the smallest values of $\delta_t/t_{\parallel}$ the degeneracy in the ES and the uniform fermion density persist, suggesting that the topological boundary modes are only destroyed at finite values of the symmetry-breaking parameter. A similar effect was observed in Ref.~\cite{Kraus}, where the authors saw signatures of topological features in the $\phi=0$ chain for small but finite interchain single-particle hopping---the relevant symmetry-breaking perturbation in that case. This may be because the spin sector is gapped and completely decoupled from the gapless charge sector; hence its behavior is robust to sufficiently small perturbations.

\subsection{Making imbalanced local potentials}

\begin{figure}[ht]
\centering
\includegraphics[width=0.75\textwidth]{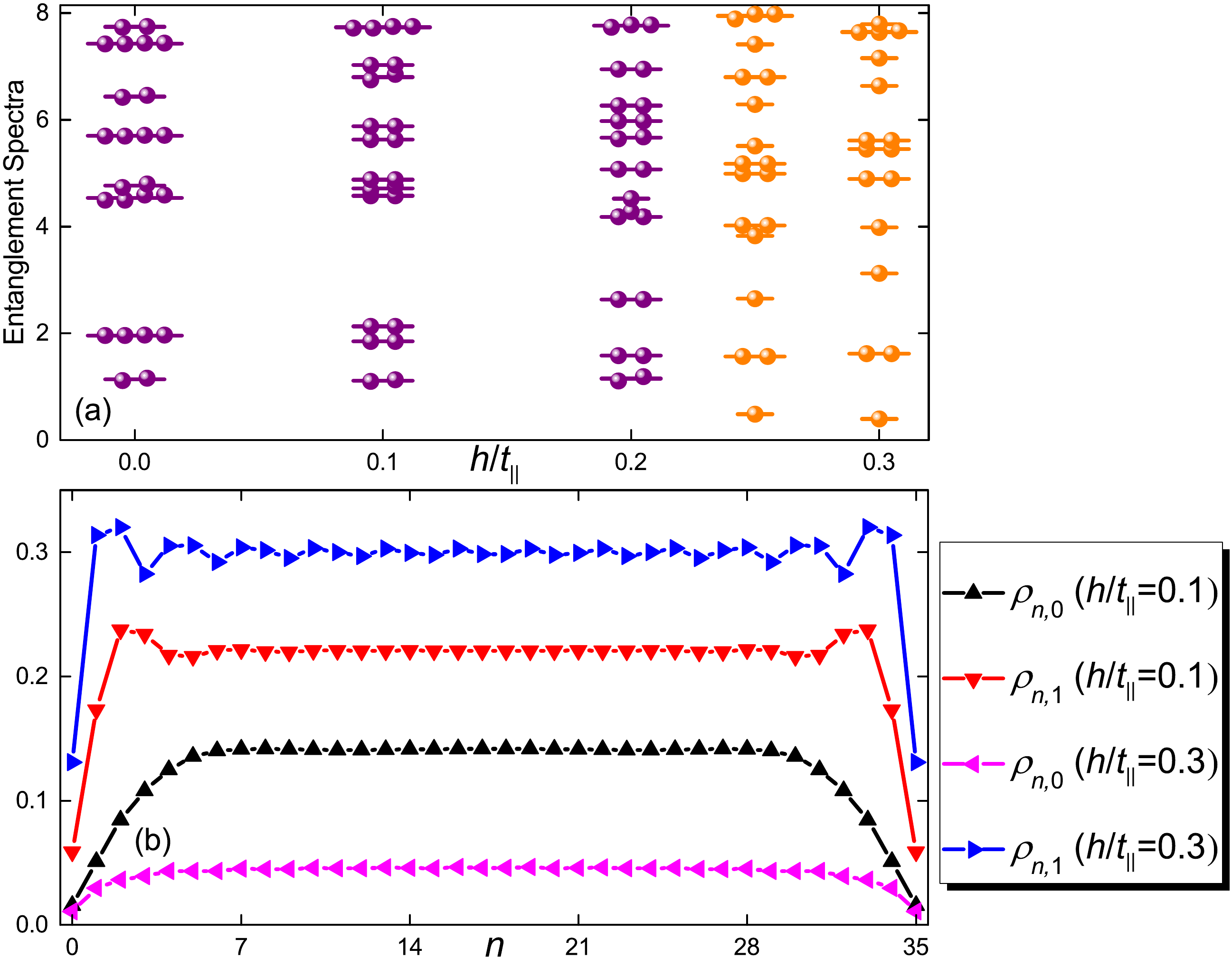}
\caption{\label{fig:fig_supp_Zeeman_den} DMRG results: (a) shows the evolution of ES (for a system of $L=36,\ N=12$) as a function of ${\sf L_s}$-breaking perturbation $h/t_{\parallel}$ (see Eq.~(\ref{Ham_Zeeman})). The even degeneracy of ES survives up to small values of $h/t_{\parallel}\approx0.225$. Panel (b) shows the development of weak density modulations along the lower chain as $h/t_{\parallel}$ increases. Here $W=-1.7t_{\parallel},\ t_{\perp}=t_{\parallel},\ \phi=\pi$.}
\end{figure}

As a second test, we consider adding a chemical potential that is imbalanced between the two chains of the Fermi ladder. We thus take the Hamiltonian to be
\begin{align}
H'=&-\sum^{L-2}_{n=0}\left[\left(it_{\parallel}  c^{\dagger}_{n,0}c_{n+1,0}-it_{\parallel} c^{\dagger}_{n,1}c_{n+1,1}\right)+\textrm{H.c.}\right] -\sum^{L-1}_{n=0}\left(t_{\perp}c^{\dagger}_{n,0}c_{n,1} +\textrm{H.c.}\right) \nonumber \\
&+\sum^{L-2}_{n=0}\left(Wc^{\dagger}_{n,0}c^{\dagger}_{n+1,0} c_{n,1}c_{n+1,1}+\textrm{H.c.}\right) +\sum^{L-1}_{n=0}\left(hc^{\dagger}_{n,0}c_{n,0}-hc^{\dagger}_{n,1}c_{n,1}\right),
\label{Ham_Zeeman}
\end{align}
where the last $h$-terms also break the ${\sf L_s}$ symmetry explicitly.

Via DMRG, we find that the effect of $h$-terms (see Fig.~\ref{fig:fig_supp_Zeeman_den}) is similar to that of the $\delta_t$-terms, both of which tend to destroy the topological Majorana boundary modes as indicated by the splitting (or nondegeneracy) in the lowest level of ES. This perturbation also has a similar effect on the fermion-density profile, first changing the relative occupancies of the two chains, and at higher values leading to density modulations in the higher density chain. The two perturbations differ quantitatively, however: In this case, the two-fold ES degeneracy survives up to $h/t_{\parallel}$ between about $0.2$ and $0.25$. The other parameters are kept the same as that in Fig.~\ref{fig:fig_supp_diff_t_parallel_den}: $W=-1.7t_{\parallel},\ t_{\perp}=t_{\parallel},\ \phi=\pi,\ L=36,\ N=12$.

\subsection{Making the flux deviate from pi}

\begin{figure}[ht]
\centering
\includegraphics[width=0.75\textwidth]{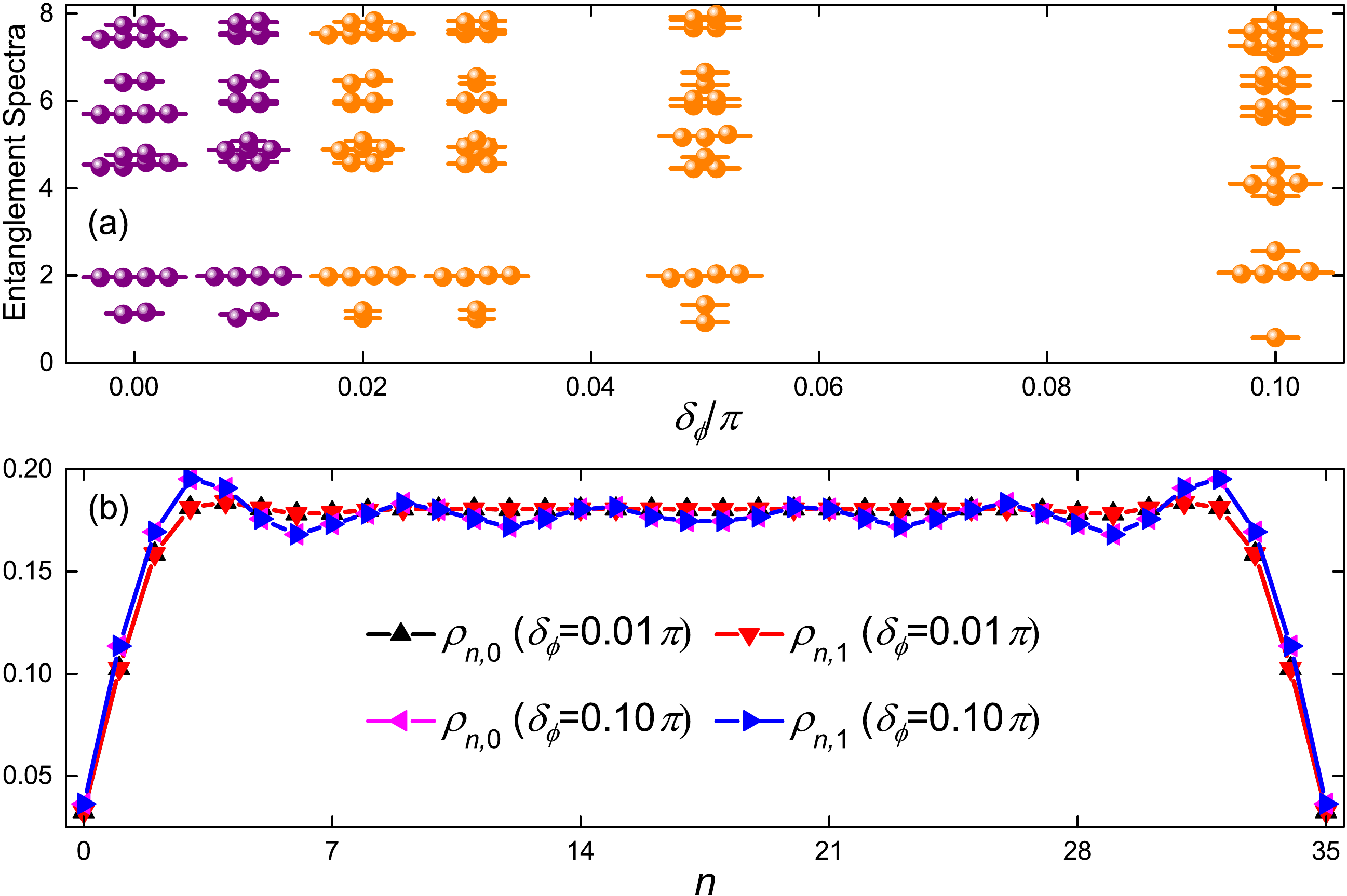}
\caption{\label{fig:fig_supp_dev_pi_den} DMRG results: (a) shows the evolution of ES (for a system of $L=36,\ N=12$) as a function of ${\sf L_s}$-breaking perturbation $\delta_{\phi}/\pi$ (see Eq.~(\ref{Ham_dev_pi})). The even degeneracy of ES survives only up to very small values of $\delta_{\phi}/\pi\lesssim0.03$. Panel (b) shows the development of a fermion-density-wave order under the increase of $\delta_{\phi}$. Here $W=-1.7t_{\parallel},\ t_{\perp}=t_{\parallel},\ \phi=\pi$.}
\end{figure}

Finally, we check numerically the consequences of the deviations of $\phi$ from the specific value of $\pi$. The corresponding Hamiltonian takes the following form,
\begin{align}
H'=&-\sum^{L-2}_{n=0}\left[\left(t_{\parallel} e^{\frac{\imi}{2}\left(\pi-\delta_{\phi}\right)} c^{\dagger}_{n,0}c_{n+1,0}+t_{\parallel} e^{-\frac{\imi}{2}\left(\pi-\delta_{\phi}\right)} c^{\dagger}_{n,1}c_{n+1,1}\right)+\textrm{H.c.}\right] -\sum^{L-1}_{n=0}\left(t_{\perp}c^{\dagger}_{n,0}c_{n,1} +\textrm{H.c.}\right) \nonumber \\
&+\sum^{L-2}_{n=0}\left(Wc^{\dagger}_{n,0}c^{\dagger}_{n+1,0} c_{n,1}c_{n+1,1}+\textrm{H.c.}\right),
\label{Ham_dev_pi}
\end{align}
where the extra $\delta_{\phi}$-phases break the ${\sf L_s}$ symmetry explicitly.

Qualitatively, our results are similar to the two cases described above: A small perturbation $\delta_\phi$ is sufficient to lift the degeneracy of the ES, signaling a loss of the topological boundary modes. However, as demonstrated in Fig.~\ref{fig:fig_supp_dev_pi_den}(a), for $t_{\perp}/t_{\parallel}=1$ there is a significant quantitative difference: The topological ES degeneracy is significantly more sensitive to the $\delta_{\phi}$-perturbations than to the $\delta_t$- and $h$-terms. The lowest-level two-fold degeneracy in the ES becomes split at a small value of $\delta_{\phi}/\pi\lesssim0.03$, indicating the instability of the Majorana boundary modes. We note, however, that this quantitative effect is necessarily sensitive to the magnitude of $t_\perp$, since for $t_\perp=0$ the flux has less impact on the band structure. Thus for small interchain single-particle tunneling strengths, the ES degeneracy survives to larger values of $\delta_{\phi}/\pi$.

As discussed in the main text, as the flux deviates from $\pi$ at finite $t_\perp$, a commensurate-incommensurate transition is expected. In this case, our numerics suggest that this transition is into a regime with a fermion density-wave-type order. This is shown in Fig.~\ref{fig:fig_supp_dev_pi_den}(b) for $W=-1.7t_{\parallel},\ t_{\perp}=t_{\parallel},\ \phi=\pi,\ L=36,\ N=12$.

\end{document}